\documentclass[aip,apl,reprint,twocolumn,numerical,superscriptaddress,amsmath,amssymb]{revtex4-1}

\usepackage[]{graphicx}
\usepackage[utf8]{inputenc}
\usepackage[T1]{fontenc}%
\usepackage{dcolumn}%
\usepackage{bm}%
\usepackage{lmodern}
\usepackage[english]{babel}
\usepackage{hyperref}
\usepackage{xcolor}
\usepackage{float}
\usepackage{footmisc}
\PassOptionsToPackage{hyphens}{url}

\newcommand{\beq}{\begin{equation}\begin{aligned}}
\newcommand{\eeq}{\end{aligned}\end{equation}}
\newcommand*{\citen}[1]{%
  \begingroup
    \romannumeral-`\x % remove space at the beginning of \setcitestyle
    \setcitestyle{numbers}%
    \cite{#1}%
  \endgroup   
}

\definecolor{linkcol}{rgb}{0,0,0.4}
\definecolor{citecol}{rgb}{0.5,0,0}

\hypersetup{
bookmarksopen=true,
pdftitle="Light-emitting field-effect transistors using Li-ion conductive glass ceramics",
pdfauthor="M. Philippi et al.",
pdfsubject="Light-emitting field-effect transistors using Li-ion conducting glass ceramics", %subject of the document
pdfstartview={FitH},    % fits the width of the page to the window
pdfmenubar=true, %menubar shown
pdfhighlight=/O, %effect of clicking on a link
colorlinks=true, %couleurs sur les liens hypertextes
pdfpagemode=UseNone, %aucun mode de page
pdfpagelayout=SinglePage, %ouverture en simple page
pdffitwindow=true, %pages ouvertes entierement dans toute la fenetre
linkcolor=linkcol, %couleur des liens hypertextes internes
citecolor=linkcol, %couleur des liens pour les citations
urlcolor=blue
}

\begin{document}
%\preprint{}

\title{Lithium-ion conducting glass ceramics for electrostatic gating}
\author{Marc Philippi}
\affiliation{DQMP and GAP, Université de Genève, 24 quai Ernest Ansermet, CH-1211, Geneva, Switzerland}
\author{Ignacio Gutiérrez-Lezama}
\affiliation{DQMP and GAP, Université de Genève, 24 quai Ernest Ansermet, CH-1211, Geneva, Switzerland}
\author{Nicolas Ubrig}
\affiliation{DQMP and GAP, Université de Genève, 24 quai Ernest Ansermet, CH-1211, Geneva, Switzerland}
\author{Alberto F. Morpurgo}
\email{alberto.morpurgo@unige.ch}
\affiliation{DQMP and GAP, Université de Genève, 24 quai Ernest Ansermet, CH-1211, Geneva, Switzerland}

\date{\today}
\begin{abstract}
We explore solid electrolytes for electrostatic gating using field-effect transistors (FETs) in which thin WSe$_2$ crystals are exfoliated and transferred onto a lithium-ion conducting glass ceramic substrate. For negative gate voltages ($V_G < 0$) the devices work equally well as ionic liquid gated FETs while offering specific advantages, whereas no transistor action is seen for $V_G>0$. For $V_G <0$ the devices can nevertheless be driven into the ambipolar injection regime by applying a large source-drain bias, and strong electroluminescence is observed when direct band-gap WSe$_2$ monolayers are used. Detecting and imaging the emitted light is much simpler in these FETs as compared to ionic liquid gated transistors, because the semiconductor surface is exposed (i.e., not covered by another material). Our results show that solid electrolytes are complementary to existing liquid gates, as they enable experiments not possible when the semiconductor is buried under the liquid itself.

\end{abstract}
%\pacs{}
%\keywords{scanning photocurrent microscopy;transition metal dichalcogenides}

\maketitle

\begin{figure}[h]
\centering
\includegraphics[width=.5\textwidth]{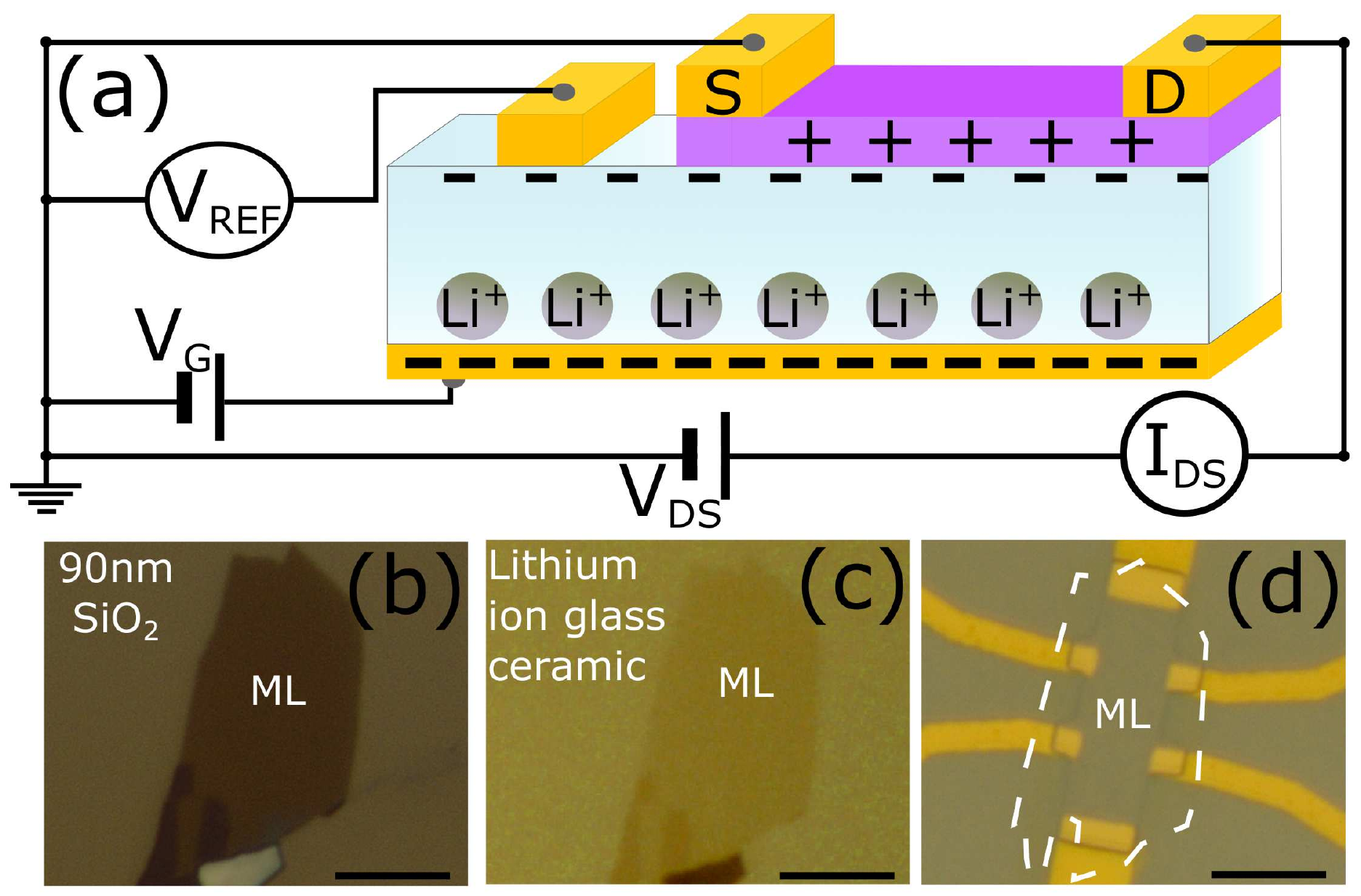}%
\hspace{0.02\textwidth}%
\caption{(a) Diagram of a ICGC-gated device, with the schematics of the electrical connections ($V_{REF}$ is the potential measured at the reference electrode). (b)	Optical microscope image of a monolayer WSe$_2$ crystal on a 90 nm SiO$_2$ substrate. (c) The same monolayer imaged in dark field after having been transferred onto an ICGC substrate. (d) Bright-field microscope image of the same monolayer after deposition of the electrical contacts (see how the monolayer, indicated by the white dashed line, is virtually invisible in bright field images). The scale bar in \textbf{b},\textbf{c} and \textbf{d} is 5 $\mu$m.}
\label{fig1}
\end{figure}

Modulating the charge carrier density at the surface of semiconductors or insulators is commonly
done by means of electrostatic gating in field-effect transistors (FETs), employing conventional solid state dielectrics. In these devices, dielectric breakdown typically limits the maximum accumulated density to approximately  10$^{13}$ cm$^{-2}$. Drastically higher carrier density values --up to 10$^{15}$ cm$^{-2}$, corresponding roughly to one electron per surface atom-- can be reached by employing ionic liquids or gels, \cite{panzer_low-voltage_2005,shimotani_electrolyte-gated_2006,shimotani_insulator--metal_2007,ueno_electric-field-induced_2008,yuan_hongtao_highdensity_2009,scherwitzl_electric-field_2010,ueno_discovery_2011,ye_accessing_2011,ye_superconducting_2012,braga_quantitative_2012,shi_superconductivity_2015}
which exploit the very large capacitance (1-50 $\mu$F/cm$^2$)\cite{yuan_hongtao_highdensity_2009} associated to the thin (nanometer) interfacial double layer.  As a result, ionic gated FETs could be used to demonstrate the occurrence of gate induced superconductivity\cite{ueno_electric-field-induced_2008,ueno_discovery_2011,jo_electrostatically_2015,shi_superconductivity_2015,costanzo_gate-induced_2016} at the surface of large gap semiconductors, to generate electroluminescence by enabling simultaneous injection of electrons and holes in different semiconducting materials\cite{jo_mono-_2014,zhang_electrically_2014,ponomarev_ambipolar_2015,gutierrez-lezama_electroluminescence_2016} and to develop new forms of spectroscopy useful to characterize 2D semiconductors \cite{braga_quantitative_2012,lezama_surface_2014,ponomarev_ambipolar_2015,xu_reconfigurable_2015,gutierrez-lezama_electroluminescence_2016,prakash_bandgap_2017}.\\ %
Despite these impressive achievements, current ionic-gated FETs suffer from  different drawbacks. An important issue is that in all devices realized so far the ionic liquid (or gel) covers the gated material, impeding the use of different experimental techniques that rely on surface sensitive probes (e.g., scanning tunneling microscopy and spectroscopy, photoemission spectroscopy, etc.). The covering liquid also adds considerable complexity to all experiments in which emission of light from the gated material is of interest, since it can absorb or deflect the light and alter the signal\cite{jo_mono-_2014}. Finally, besides practical problems such as the very strong sensitivity to air and humidity\cite{capelli_organic_2010}, ionic liquids also pose constraint on device fabrication and on the type of structures that can be realized, because it is virtually impossible to perform any additional nano-fabrication step after deposition of the liquid itself.\\
To address these problems we started exploring the possibility to substitute the liquid/gel electrolytes that have been employed until now with solid electrolytes. To this end, we have identified as promising candidates ion-conductive glass ceramics (ICGCs)\cite{xu_lithium_2007,knauth_inorganic_2009,hayashi_superionic_2012}, in which Li$^+$ or Na$^+$ ions are free to move within a solid framework composed of different oxide materials. ICGCs are used as components in batteries (separators, electrolytes and cathodes), are known to have a high ionic conductivity (10$^{-3}$ - 10$^{-4}$ Scm$^{-2}$)\cite{xu_lithium_2007,knauth_inorganic_2009,hayashi_superionic_2012}, and to be stable under ambient conditions. The strategy that we follow to realize a FET device is to deposit a layer of semiconductor on top of a Li-ICGC substrate with a back metallic layer acting as a gate and micro-fabricated contacts enabling transport measurements (see scheme in Fig. 1a). In such a configuration, device operation is conceptually  similar to that of conventional liquid electrolyte devices: upon application of a gate voltage, the Li$^+$ or Na$^+$ ions in the glass matrix move creating a space charge region near the surface due to depletion or accumulation of positively charged ions (depending on the gate voltage polarity), allowing the potential to be transferred from the gate electrode to the device.\\
Despite the conceptual similarity, it is essential to investigate the detailed aspects of the device response, as in practice they may differ significantly from the case of ionic-liquid gated FETs. For instance, it remains to be seen whether in solid electrolytes the spatial extension of the space charge region\cite{paradisi_space_2015} at the surface is sufficiently small to generate the large required electrical capacitance. It is also unclear whether devices can function properly for both polarities of applied gate voltage. Indeed, while for a negative gate voltage the ions are attracted to the gate and no problem is expected, for a positive applied gate voltage ions are pushed toward the device. As Na$^+$ --and even more Li$^+$-- ions are known to be very mobile, they may cause unwanted effects of none-electrostatic nature (e.g., intercalation, chemical reactions, etc.)\cite{hermann_electrical_1973,woollam_superconducting_1976,biscaras_onset_2015,yu_gate-tunable_2015,sterpetti_comprehensive_2017}, thereby preventing appropriate transistor operation.\\
To test the operation of Li-ICGC FETs we have realized devices based on thin exfoliated semiconducting transition metal dichalcogenides (TMDs), whose well-studied behavior in ionic liquid FETs provides an excellent reference\cite{chuang_high_2014,pradhan_hall_2015,wang_electronic_2015,prakash_bandgap_2017,pudasaini_high_2017} (most of the work discussed here is based on devices realized with WSe$_2$ crystals purchased from 2D semiconductors).  We find that upon the application of a negative gate voltage electrostatic doping works properly and allows reaching hole densities as high as 7 x 10$^{13}$ cm$^{-2}$, comparable to values achieved by ionic-liquid gating on several TMDs\cite{braga_quantitative_2012,lezama_surface_2014,zhang_ambipolar_2012} (the largest accumulated hole density in common semiconducting TMDs is typically smaller than the largest possible electron density\cite{braga_quantitative_2012}). For this gate voltage polarity, virtually all our devices exhibit high-quality transistor characteristics. Upon the application of positive gate voltage, however, no FET action is seen, confirming that effects other than simple electrostatics play a role when alkali ions are pushed to the surface. It is nevertheless possible to bias the devices into the ambipolar injection regime (i.e., with electrons and holes simultaneously injected at opposite contacts) by applying a negative gate bias concomitantly with a larger negative source-drain bias. In this regime, when a direct-gap monolayer WSe$_2$ is used as transistor channel, electro-luminescence from exciton recombination is clearly observed. We find that controlling the induced electroluminescence is much easier than in ionic-liquid gated FETs, because the transistor channel is directly exposed to air and not buried under the ionic liquid/gel itself. Overall, our results demonstrate that Li$^+$-ICGCs substrates can be implemented as gate dielectrics in actual devices with a number of advantages over conventional ionic-liquid/gel gated transistors. \\
The fabrication of WSe$_2$ devices on Li-ICGC substrates consists of several steps. Thin WSe$_2$ crystals are exfoliated onto Si/SiO$_2$ substrates (see Fig 1b), and subsequently transferred onto commercially available 150 $\mu$m thick ICGCs substrates purchased from MTI corporation (see Ref. \citen{mtixtl_conductive_nodate} for the material data sheets and additional information), via a pick-up and release technique based on PC/PDMS stacks\cite{zomer_fast_2014}. This proved necessary because even relatively thick (e.g., 10 nm) exfoliated crystals exhibit extremely small contrast on ICGCs substrates and are very difficult to identify unless dark field imaging is employed (see Fig. \ref{fig1}c and Fig. \ref{fig1}d, which compare dark and bright field optical microscope images of a same  WSe$_2$ monolayer). Electrical connections to the transferred crystals --as well as a reference electrode on top of the ICGC substrate-- are then realized by means of fully conventional electron beam lithography, evaporation of Pt/Au films and lift-off. Fig. 1d shows an optical microscope image of a complete monolayer WSe$_2$ device. Electrical measurements on more than ten devices realized with monolayer, bilayer and thicker crystals, all showing the same transistor behavior, were performed using a SR830 lock-in amplifier, Keithley 2400 source/measure unit,  Agilent 34401A digital multimeters, and home-made voltage/current amplifiers. In some cases, measurements were performed with an  Agilent Technology E5270B parameter analyzer, which gave identical results.\\
\\
We start the characterization of  Li-ICGC gated devices by measuring the FET transfer curve (source-drain current $I_{DS}$ as a function of gate voltage $V_G$) for negative  $V_G$. Fig. 2a shows the current in the linear regime for different values of applied source-drain voltage $V_{DS}$ (=1, 4, and 10 mV). The current $I_{DS}$ scales linearly with $V_{DS}$, as shown by the fact that measurements at different $V_{DS}$ lead to the same square conductance (see Fig. 2b; the identical conductivity found in successive measurements also illustrates the excellent reproducibility and stability of the devices). The threshold voltage for hole conduction is $V_{th} = -2.1$ V, after which the current measured for $V_{DS}=10$ mV rapidly reaches hundreds of nA, much larger than the leakage current typically measured (of the order of 1 nA or less, see Fig. 2c). The measured ON/OFF ratio ($10^4$; see Fig. 2d) is determined by the current noise floor of our experimental set-up. From these measurements the maximum hole density accumulated at $V_G = -3$ V is estimated by comparing the conductivity to the one measured on ionic-liquid gated WSe$_2$\cite{prakash_bandgap_2017} and other semiconducting TMD FETs \cite{braga_quantitative_2012,lezama_surface_2014,zhang_ambipolar_2012}, at a same difference between threshold and applied gate voltage. The conductivity values in Li-ICGC and ionic-liquid gated devices are comparable, and using a characteristic hole mobility value of  $\mu_h \simeq 50$ cm$^2$/Vs \cite{braga_quantitative_2012,lezama_surface_2014,zhang_ambipolar_2012} we estimate the largest density of accumulated holes to be 7 x 10$^{13}$ cm$^2$/Vs. These results, obtained on a transistor realized on an exfoliated WSe$_2$ crystal that is approximately 15 nm thick, are representative of what we commonly observed in many different Li-ICGC transistors that we investigated.\\
Other aspects of the data point to the quality of Li-ICGC FETs. One is the value of the sub-threshold swings $S$, extracted from the logarithmic plot of $I_{DS}$-vs-$V_G$ (See Fig. 2d). We obtain $S\simeq 80$ mV/dec, fully comparable to the values found in ionic-liquid gated FETs\cite{braga_quantitative_2012,lezama_surface_2014,xu_reconfigurable_2015}, and very close to the ultimate room temperature limit of 60 mV/decade. Finding that $S$ approaches 60 mV/decade provides a direct confirmation of the large capacitance\cite{m._sze_physics_2006,braga_quantitative_2012} of the glass ceramic gates. Another aspect is the virtually complete absence of hysteresis upon sweeping $V_G$ up to large negative voltages and back, which was observed in essentially all devices for sweeping rates up to $\approx 5$ mV/s upon repeatedly cycling the gate voltage. In contrast, in ionic-liquid gated FETs a somewhat larger hysteresis is often found when operating the devices under identical conditions. Finally, in nearly all Li-ICGC gated FETs investigated, the efficiency of the gate --defined as the ratio between the voltage $V_{REF}$ measured at the reference electrode and the applied gate voltage $V_G$ -- is reproducibly close to 100\% (see Fig. 2e), indicating that the applied V$_G$ drops almost entirely across the ceramic/channel interface. This is consistently better than the 60-80\% efficiency typically observed in ionic-liquid gated devices (see, for instance, reference 17).\\
\begin{figure}[h]
\centering
\includegraphics[width=.5\textwidth]{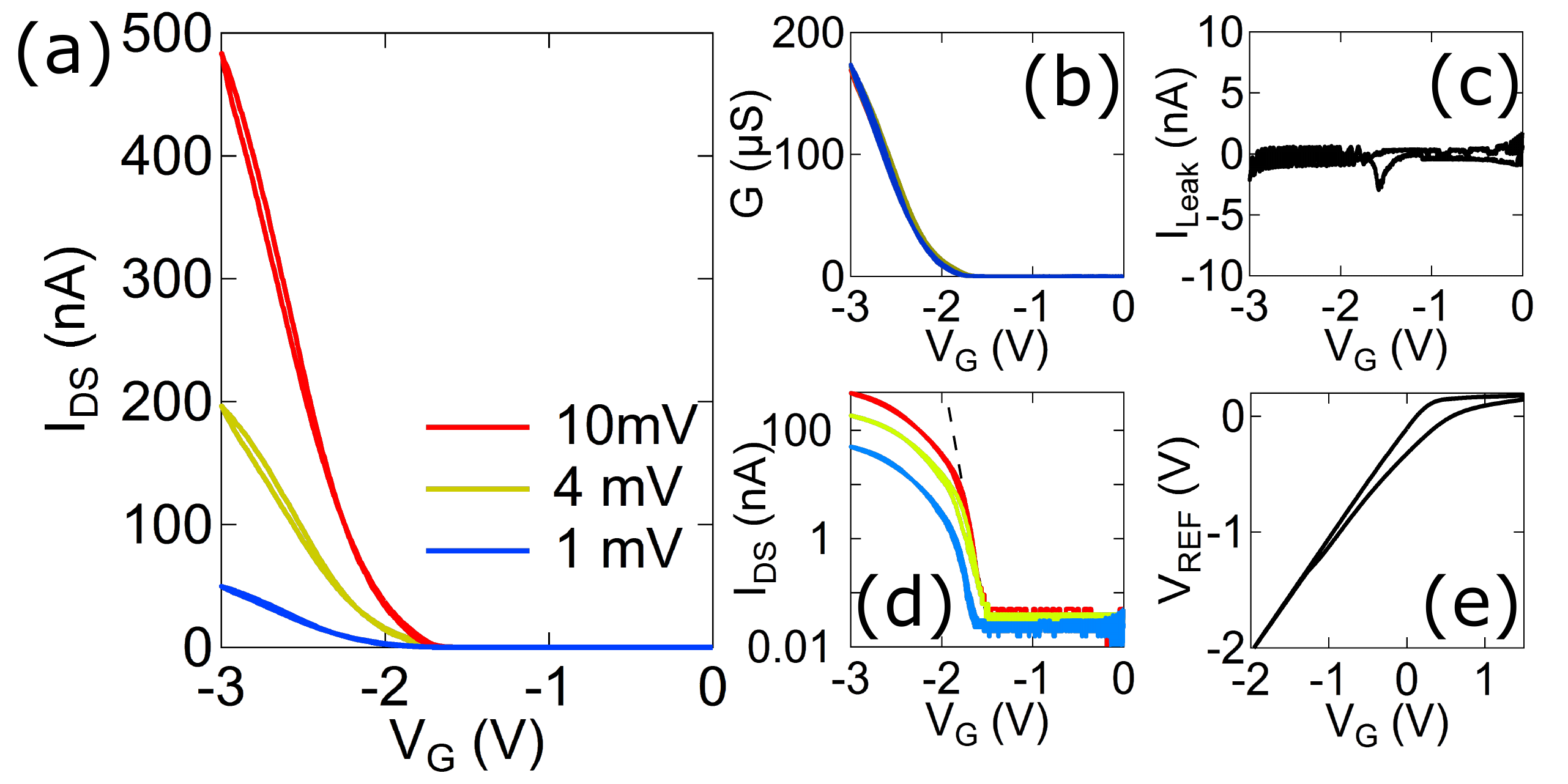}%
\hspace{0.02\textwidth}%
\caption{(a) $I_{DS}$ vs $V_G$ curves (transfer curves) of a device realized on an approximately 15 nm thick WSe$_2$ exfoliated crystal, measured at room temperature for $V_{DS} =$ 1 mV, 4 mV and 10 mV. (b) shows the same data plotted in terms of the square conductance. (c) Leakage current corresponding to the data of (a) measured as a function of $V_G$. (d) Transfer curves shown in (a) plotted in a logarithmic scale to extract the subthreshold slope. (e) The reference potential $V_{REF}$ as a function of $V_G$ exhibits systematically an efficiency close to 100\% for  $V_G < 0$ and saturation at small values for $V_G > 0$.}
\label{fig2}
\end{figure}\\
It follows from these considerations that --under a negative applied gate voltage-- the performance of our Li-ICGC gated devices is fully comparable to that of ionic-liquid gated FETs, and even superior in a number of regards. The situation is however very different under the application of a positive gate voltage, in which case no transistor action was ever observed in our measurements. That something unusual happens in the Li-ICGC devices for $V_G >0$ is clearly apparent from the plot of $V_{REF}$-vs-$V_{G}$ in Fig. 2e, in which we see that $V_{REF}$ does not increase upon increasing $V_G$, but remains pinned at a small positive value. This behaviour clearly indicates that upon the application of a positive $V_G$, the action of the gate is not determined by simple electrostatics:  some other phenomenon likely related to the accumulation of a large density of Li ions at the surface (possibly the formation of a metallic Lithium layer sufficient to screen the applied gate voltage) prevents the devices to be operated in electron accumulation mode.\\
We now proceed to analyze the device output characteristics ($I_{DS}$-vs-$V_{DS}$ measured at fixed, negative $V_G$). Fig. 3a shows data obtained from the same device whose transfer curves are shown in Fig. 2a. The expected transistor behaviour is apparent, with the current $I_{DS}$ initially increasing linearly at small $V_{DS}$ and exhibiting full saturation due to pinch-off when $|V_{DS}| > |V_G-V_{th}|$. Upon increasing $V_{DS}$ well past the onset of saturation, a sharp increase of $I_{DS}$ is seen in all devices investigated. This is shown in Fig. 3b, with data measured on a Li-ICGC device realized with a WSe$_2$ monolayer. Such an increase occurs when  $V_{DS}$ is sufficiently large to invert the potential of the semiconducting channel at the drain contact. When that happens, charge carriers of opposite polarity are injected by the source and drain electrodes and the current is carried simultaneously by electrons and holes that recombine inside the channel. This is the so-called ambipolar injection regime\cite{kang_moon_sung_pedagogical_2013} commonly observed in ionic-liquid gated FETs based on most semiconducting TMDs\cite{braga_quantitative_2012,zhang_ambipolar_2012,zhang_formation_2013,lezama_surface_2014,chuang_high_2014}. In Li-ICGC gated transistors, detecting the occurrence of this regime is particularly interesting, because the absence of any current upon the application of a positive $V_G$ might have suggested that electron conduction may not happen. The data in Fig. 3b shows that this is not the case.\\
\begin{figure}[h]
\centering
\includegraphics[width=.5\textwidth]{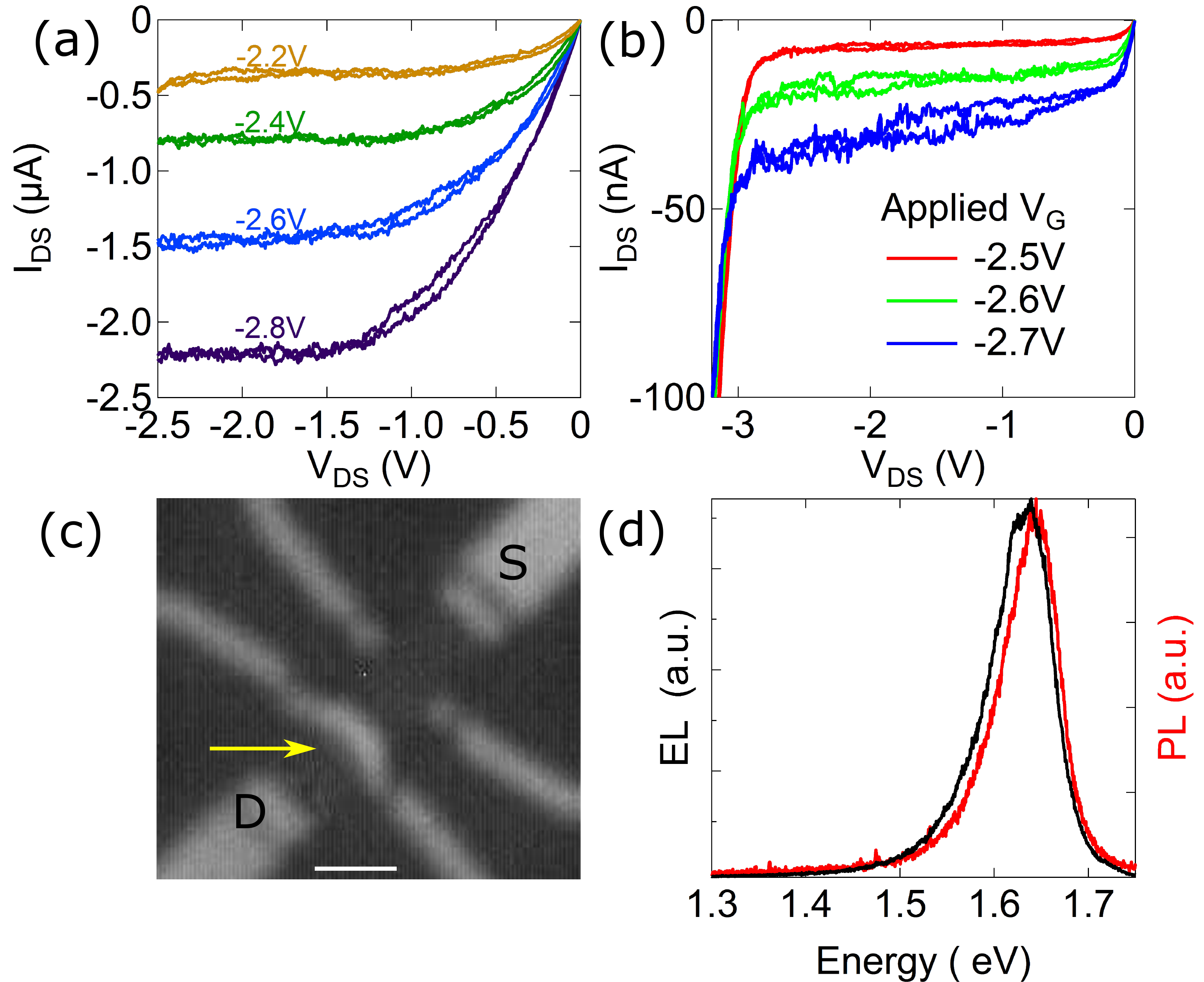}%
\hspace{0.02\textwidth}%)
\caption{(a) $I_{DS}$ vs $V_{DS}$ curves (output curves) of the same device whose transfer curves are shown in Fig \ref{fig2}, for $V_G =$ -2.2 V, -2.4 V, -2.6 V and -2.8 V. (b) Ambipolar injection regime reached in the output curves of a monolayer WSe$_2$ device at different $V_G$ values (see legend) (c) Microscope image of a monolayer WSe$_2$ device biased in the ambipolar regime ($V_G$ = -2.5 V and $V_{DS}$ = -5.1 V). The yellow arrow points to the emitted light (S and D label the source and drain contacts). (d) Comparison between the electroluminescence (EL; black curve) and photoluminescence (PL; red curve) spectra measured on the same monolayer WSe$_2$ device.}
\label{fig3}
\end{figure}
To obtain a compelling confirmation that electrons and holes are simultaneously carrying current we search for electroluminescence, which is expected in the ambipolar injection regime if the semiconductor has a direct band-gap. This is the case for monolayers WSe$_2$ and we indeed find that --in the ambipolar injection regime-- current flow is accompanied by light emission. To detect the emitted light we mounted our WSe$_2$ monolayer device inside a vacuum chamber with optical access under an optical microscope, and used a common digital camera to take images (see Fig. 3c). Light is emitted rather uniformly across the entire device, at the position where electrons and holes meet. To identify the process responsible for light emission we measured the spectral dependence of the intensity using an Andor Shamrock spectrometer. We find that the electroluminescence spectrum coincides nearly perfectly with the spectral dependence of monolayer WSe$_2$ photoluminescence, as shown in Fig. 3d, which allows us to conclude that electroluminescence is due to recombination of direct excitons (photoluminescence was measured using a Fianium Supercontinuum laser coupled to a monochromator at  $\lambda$ = 610 nm at a power of 10 $\mu$W to excite the WSe$_2$ monolayer). We remark that the use of the glass ceramic gate allows for a direct comparison of both spectra without having to subtract a background, as we found necessary to do in similar ionic-liquid gated devices based on WS$_2$ monolayers\cite{jo_mono-_2014}. This shows that the optical signal originating from the device is stronger and cleaner as compared to the case of ionic-liquid gated devices, because on Li-ICGC FETs no other material is present on top.\\
Finally, we discuss the evolution of the emitted light upon increasing $V_{DS}$. Fig 4a-d show microscope images of a monolayer WSe$_2$ FET as $V_{DS}$ is increased from -3 V to -6 V ($V_G = -2.5$ V in all cases). For $V_{DS}=-3$ V (Fig. 4a) the device is very close to the onset of the ambipolar injection regime (see Fig. 3b) and no light emission is detected yet. At $V_{DS}=-4.5$ V (Fig. 4b) light emission starts to be clearly visible close to the drain electrode. A further increase in $V_{DS}$ ($V_{DS} = -5.1$ V in Fig. 4c and $V_{DS} = -6$ V  in Fig. 4d) causes light emission to be progressively shifted inside the transistor channel, away from the drain and towards the source contact. For all investigated bias values, light emission occurs from a line across the channel, approximately perpendicular to the current path.  This behavior originates from the presence of a pn junction in the FET channel \cite{zaumseil_spatial_2006,bisri_high_2009,kang_moon_sung_pedagogical_2013,jo_mono-_2014,zhang_formation_2013}, whose initial formation takes place near the contact where the local channel potential is reversed first due to the application of a large and negative source-drain bias. Increasing the bias leads to a shift of the position of the pn junction that follows from simple electrostatic considerations (see, e.g., Ref. \citen{kang_moon_sung_pedagogical_2013}). Note that the observation of such an ideal evolution eliminates any possible scenarios responsible for light emission other than the formation of a pn junction inside the channel, such as the electrons being injected from the contact due to hot carrier effects caused by the high lateral electric fields\cite{li_electric-field-induced_2015}. It is also worth remarking that such a clear evolution of the position of light emission could not be observed in similar ionic-liquid gated devices that we studied earlier based on WS$_2$ monolayers\cite{jo_mono-_2014}, because the presence of the ionic liquid itself both affected the device and decreased the imaging resolution. \\
\begin{figure}[h]
\centering
\includegraphics[width=.5\textwidth]{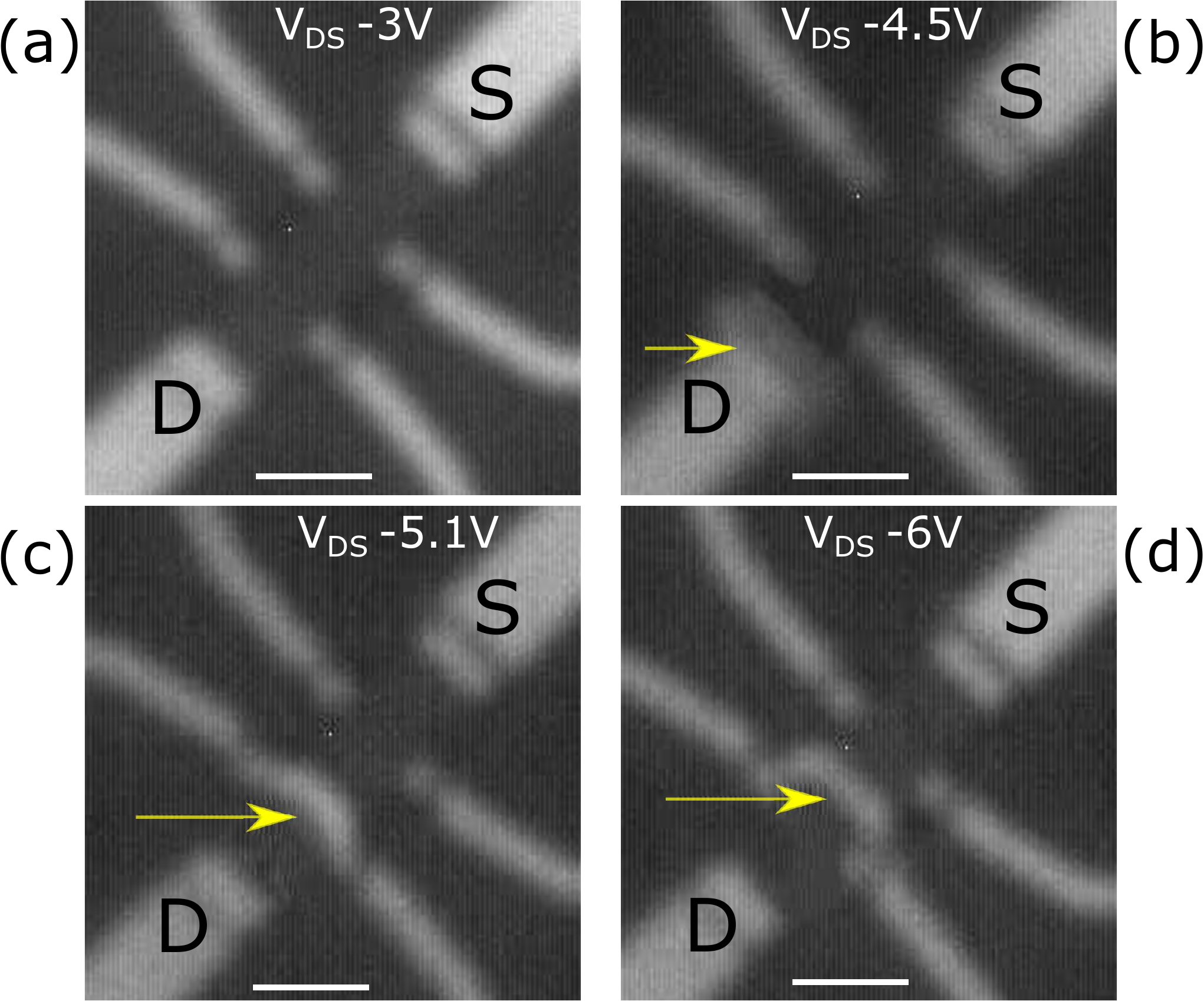}%
\hspace{0.02\textwidth}%
\caption{Microscope images of a monolayer WSe$_2$ device biased in the ambipolar regime at $V_G$= -2.5 V and different values of source-drain voltage, $V_{DS}$ = -3 V (a), $V_{DS}$ = -4.5 V (b), $V_{DS}$ = -5.1 V (c) and $V_{DS}$ = -6 V (d). (a) corresponds to the onset of the ambipolar injection regime and no light emission is seen. For larger negative $V_{DS}$ the emitted light is clearly visible and progressively shifts from the drain contact (D) region to the source contact ((S); see the yellow arrows). The scale bar is $5\mu m$ in all panels.}.
\label{fig4}
\end{figure}
In conclusion, we have demonstrated the possibility to use Li-ICGC substrates as gate dielectrics to realize transistors enabling the accumulation of a high density of charge carriers. We find that upon the application of negative gate voltage, Li-ICGC gated devices exhibit  performances comparable to that of ionic-liquid gated FETs, and offer several advantages due to the fact that the semiconductor is not buried under the liquid itself. For positive gate voltage, instead, the devices do not operate as conventional transistors, probably due to accumulation of lithium at the interface with the semiconductors. The possibility to accumulate large density of carriers (holes for the case of Li ion conducting glass ceramic) without preventing access to the material surface will be especially important in the future in the field of 2D materials, for instance to apply different surface sensitive techniques to back-gated atomically thin crystals.\\

\begin{acknowledgments}
We gratefully acknowledge A. Ferreira for technical assistance, as well as financial support from the Swiss National
Science Foundation and from the EU Graphene Flagship. NU also acknowledges funding from the Ambizione grant of the Swiss National Science Foundation. AFM acknowledges useful discussions with Prof. Yuanbo Zhang.  \\
\end{acknowledgments}
\end{document}